\documentclass[a4paper,11pt]{article}

\usepackage[latin1]{inputenc}
\usepackage[T1]{fontenc}
\usepackage{textcomp,graphicx,epsfig}
\usepackage{latexsym,amssymb,amsmath,amsthm}
\usepackage{fancyhdr,rotating,makeidx,eurosym}
\usepackage{lmodern,url,xspace,hyperref,layout}

\def\PP{\mathbb{P}}

\def\EE{\mathbb{E}}
\def\II{\mathbb{I}}
\def\dd{\mathrm{d}}
\def\ee{\mathrm{e}}

\begin{document}

\title{Some mathematical tools for the Lenski experiment}

\author{Bernard Ycart\footnote{LJK, CNRS UMR 5224,
Univ. Grenoble-Alpes, France \texttt{Bernard.Ycart@imag.fr}}%
       \and 
Agn\`es Hamon\footnote{LJK, CNRS UMR 5224,
Univ. Grenoble-Alpes, France %
         \texttt{Agnes.Hamon@imag.fr}}%
       \and 
J. Gaff\'e\footnote{LAPM,  CNRS UMR 5163, Univ. Grenoble-Alpes, France %
         \texttt{Joel.Gaffe@ujf-grenoble.fr}}%
       \and 
D. Schneider\footnote{LAPM,  CNRS UMR 5163, Univ. Grenoble-Alpes, France %
         \texttt{Dominique.Schneider@ujf-grenoble.fr}}}%

\date{\vspace{-5ex}}

\maketitle

\abstract{The Lenski experiment is a long term daily reproduction
of \emph{Escherichia coli}, that has evidenced phenotypic and genetic
evolutions along the years. Some mathematical models, that could be
usefull in understanding 
the results of that experiment, are reviewed here: stochastic and
deterministic growth, mutation appearance and fixation, competition of species.}



\section{Introduction}
This is primarily intended as a review of a very small part 
of the vast litterature on mathematical models applied to population
biology. The application that we have in mind is the \emph{Lenski
  experiment} (referred to as LE hereafter, see 
\cite{Barricketal09,Khanetal11,Philippeetal07,Schneideretal00,Schneideretal02,Staneketal09}). Here
is a much simplified description (see Philippe \emph{et al.}
\cite{Philippeetal07} for a more detailed account). 
The ancestral strain was  \emph{Escherichia Coli B}. Twelve
populations have been propagated by daily serial transfer, using a
1:100 dilution, in the same defined environment, for more than 40,000
generations. The growth medium supports a maximum of $5\times 10^7$ cells
per mL; therefore the number of cells in each one of the twelve 10~mL
vessels varies daily from $5\times 10^6$ to $5\times 10^8$. Our main focus will
be on modelling the appearance and fixation of mutations along successive
generations. The total number of mutational events that have happened
during the 40,000 generations in each population is estimated at
$3\times 10^9$ (\cite{Philippeetal07}, p.~849). Our main question here is:
how did some of the mutant strains survive and eventually
invade the population? We argue that only beneficial mutations
(\emph{i.e.} increasing the fitness) can have survived the experimental
process. Indeed when a nonbeneficial mutation occurs a given day, the
proportion of mutants in the population remains extremely small, and
therefore the chances that mutants survive the 1:100 sampling to the
next day are very small. On the contrary, a strain carrying a
beneficial mutation survives
if mutant cells can be found in sizeable amounts at the moment of
dilution. Provided the new strain survives the first day dilution,
chances are that its proportion in the global population will
rapidly increase. When that proportion is close to one, the initial
strain has a high probability to be wiped out by daily dilution.
Our objective here is to
provide a quantified basis to the above assertions. 
\vskip 2mm
Although our emphasis will mainly be on probability, we shall
use some elementary population dynamics, on which our basic
references are Murray \cite{Murray89} and Kot \cite{Kot01}. 
The stochastic tools used here, essentially discrete sampling and
birth-and-death processes, are presented in many excellent textbooks. For
probability theory, Feller's treatise (in
particular the second volume \cite{Feller71}) remains a good
reference. Ross \cite{Ross10} and Tuckwell \cite{Tuckwell95} are
oriented towards applications in biology. The basic theory of 
continuous time Markov processes is developed in
Bharucha-Reid \cite{Bharucha60} and \c{C}inlar
\cite{Cinlar75}. Ethier \& Kurtz \cite{EthierKurtz05} give a more
advanced treatment, in particular of asymptotics and convergence to
diffusion processes. Some textbooks, such as Gardiner
\cite{Gardiner04} are particularly oriented to applications. 
Being one of the founders of the theory,
Bartlett \cite{Bartlett60,Bartlett66} emphasizes 
population modelling. Moran \cite{Moran62} develops
genetic applications. Some more recent textbooks have been devoted 
to stochastic modelling in biology, such as 
Ewens \cite{Ewens04} and 
Wilkinson \cite{Wilkinson06}; 
Allen \cite{Allen03} and Durrett \cite{Durrett08} are 
of particular interest to us. 
The course given on ``Stochastic Population Systems'' 
by D. Dawson to the Summer School in probability at
PIMS-UBC in 2009 covers all the material that we shall use and much
more. The lecture notes are available on 
line\footnote{http://www.math.ubc.ca/~db5d/SummerSchool09/LectureNotes.html}. 
\vskip 2mm
The paper is organized as follows. The daily
sampling process will be examined first, in section
\ref{sampling}. Next, we shall review in section \ref{volterra} 
some deterministic models for
the evolution of two competing populations (normal and mutant cells)
in a constrained environment. On a very simple model, 
it will shown that the final proportion of mutants (\emph{i.e.} 
before the next dilution) can be
deduced from the initial proportion (\emph{i.e.} right after the
previous dilution) and the two division rates of normal and mutant
cells. Section \ref{birth} will  be devoted to the stochastic
counterpart of the deterministic models examined in section
\ref{volterra}. The emphasis will be on the convergence of stochastic to
deterministic models.
Stochastic mutation models will be reviewed in section
\ref{mutation}. There we shall focus on the first appearance of
mutations and the proportion of mutants at the end of the first day.
All models considered here are based on several parameters, among
which at least individual division rates and mutation probabilities. 
The question of estimating those parameters is therefore crucial, 
and it will be reviewed in section \ref{estimation}. 
\section{Daily sampling}
\label{sampling}
The problem of beneficial mutations being lost in populations with
periodic bottlenecks has been studied by Wahl and Gerrish \cite{WahlGerrish01}.
Our aim here is to give a simple quantitative evaluation of the survival of
mutations from one day to the next. Recall that the maximal capacity
of the medium is $n_f=5\times 10^8$. 
At the end of each day, $n_f$ cells are
present on the population, out of which $n_0=n_f/100$ are sampled. 
Assume that at the end of one day, $N$
normal cells and $M=n_f-N$ mutant cells are present. How many mutant cells
will remain after the 1:100 dilution? The simplest model assumes 
equiprobability:
assuming that the medium is homogeneous, each $n_0$-sized sample has
the same probability to be chosen for the next day. Therefore the
number of mutants surviving dilution follows the
hypergeometric distribution with parameters $n_f$, $M$ and $n_0$. In
other words, the chances to find exactly $n$ normal cells and 
$m=n_0-n$ mutant cells after dilution are:
$$
\frac{\binom{M}{m}\binom{N}{n}}{\binom{n_f}{n_0}}\;.
$$
Since $n_f$ is large, and $n_0$ is small compared to $n_f$, the
hypergeometric distribution can be correctly approximated by the
binomial with parameters $n_0$ and $p=M/n_f$ (proportion of
mutants). As a consequence, the probability to find exactly $m$ mutants is 
$$
\frac{\binom{M}{m}\binom{N}{n}}{\binom{n_f}{n_0}}
\simeq
\binom{n_0}{m}p^m(1-p)^{n_0-n}\;.
$$
If $p$ is large enough (say $p>10^{-4}$), the binomial distribution
above can be approximated by a Gaussian, with expectation $n_0p$ and
variance $n_0p(1-p)$. In other words, the proportion $p'=m/n_0$ will  be
close to $p$, a 95\% fluctuation interval being:
$$
p \pm 1.96 \frac{\sqrt{p(1-p)}}{\sqrt{n_0}}\;.
$$ 
Since $n_0= 5\,10^6$, one can express it differently by saying that
with 98.73\% probability, $p'$ is at distance at most $5\times
10^{-4}$ of $p$.
\vskip 2mm
However, if $p$ is too small, the approximation above is no longer
valid. A better approximation is given by the Poisson distribution
with parameter $n_0p$:
$$
\frac{\binom{M}{m}\binom{N}{n}}{\binom{n_f}{n_0}}
\simeq
\ee^{-n_0p}\frac{(n_0p)^m}{m!}\;.
$$
\vskip 2mm
We should like to emphasize here that when $M$ is small (a few
units), chances that all mutant cells disappear after dilution are
high. Indeed, the probability that no mutant cell remains is:
$$
\frac{\binom{N}{n_0}}{\binom{n_f}{n_0}}
\simeq
\ee^{-n_0 M/n_f}\simeq 1-\frac{M}{100}\;.
$$
For instance, if $M=10$ mutant cells are present, there is a 90\%
probability that they will all disappear after dilution. From the
100-fold increase, one can evaluate the daily number of generations from a
given cell to be of order $6$ ($2^6=64$) to $7$ ($2^7=128$). The table
below gives the values of the probability of disappearance for each
value of $M=2^d$, $d$ (the number of generations) ranging from $0$ to
$8$.
Numerical calculations (from the exact hypergeometric distribution)
were obtained through $R$ \cite{R}.
\begin{center}
\begin{tabular}{|l|ccccccccc|}
\hline
divisions&0&1&2&3&4&5&6&7&8\\\hline
cells&1&2&4&8&16&32&64&128&256\\\hline
proba of disappearance&0.99&0.98&0.96&0.92&0.85&0.72&0.53&0.28&0.08\\\hline
\end{tabular}
\end{center}
\section{Deterministic growth models}
\label{volterra}
\begin{figure}[!ht]
\centerline{
\includegraphics[width=10cm]{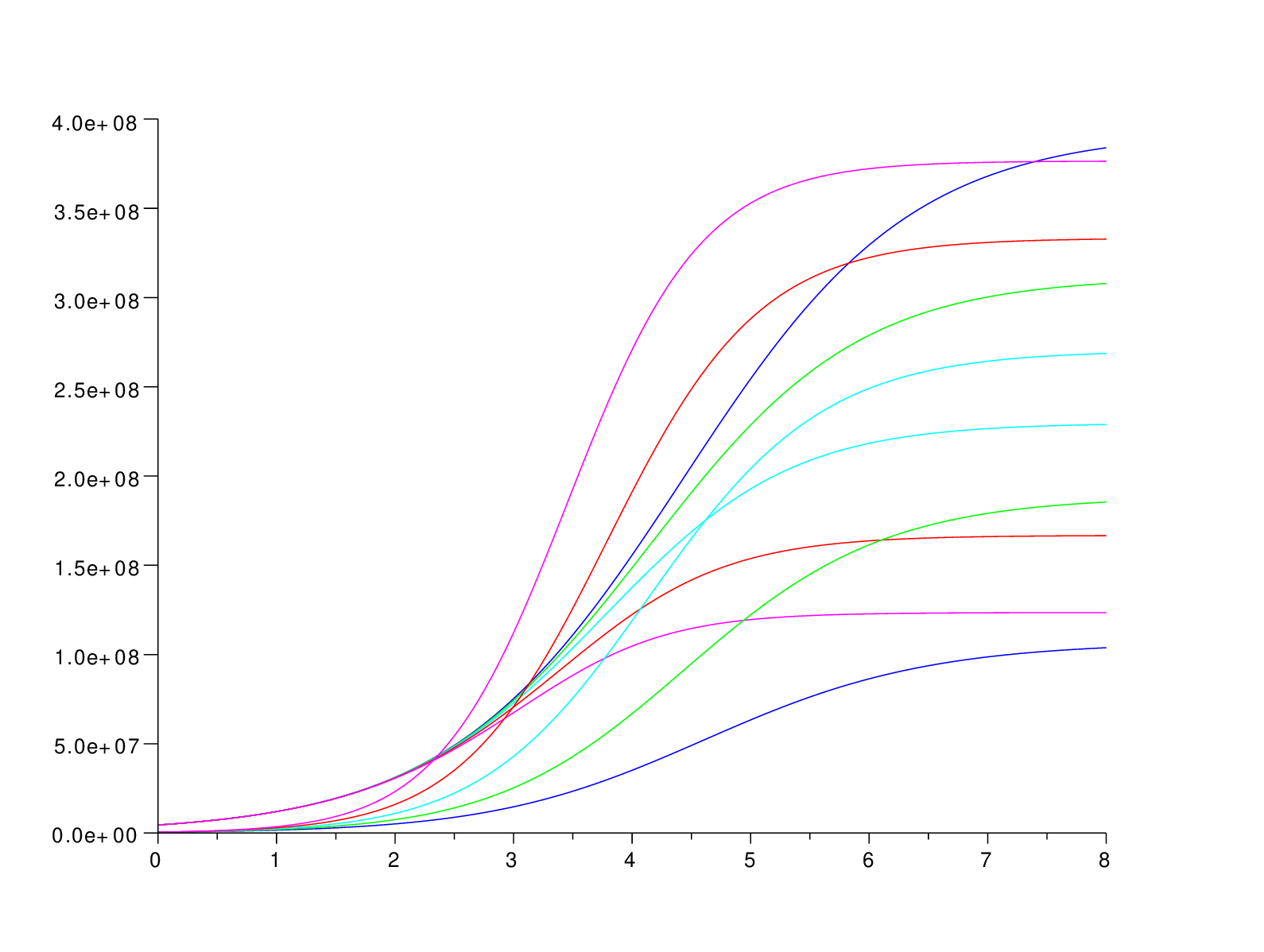}
} 
\caption{Evolution in time of the numbers of normal and mutant cells
  in time in the Volterra model. The initial proportion of mutants is
  $0.1$. The IDR of normal cells is $1$. That of
  mutant cells is $1.2$ (blue), $1.4$ (green), $1.6$ (light blue),
  $1.8$ (red), $2$ (purple).}
\label{fig:VMtraj}
\end{figure}

\begin{figure}[!ht]
\centerline{
\includegraphics[width=10cm]{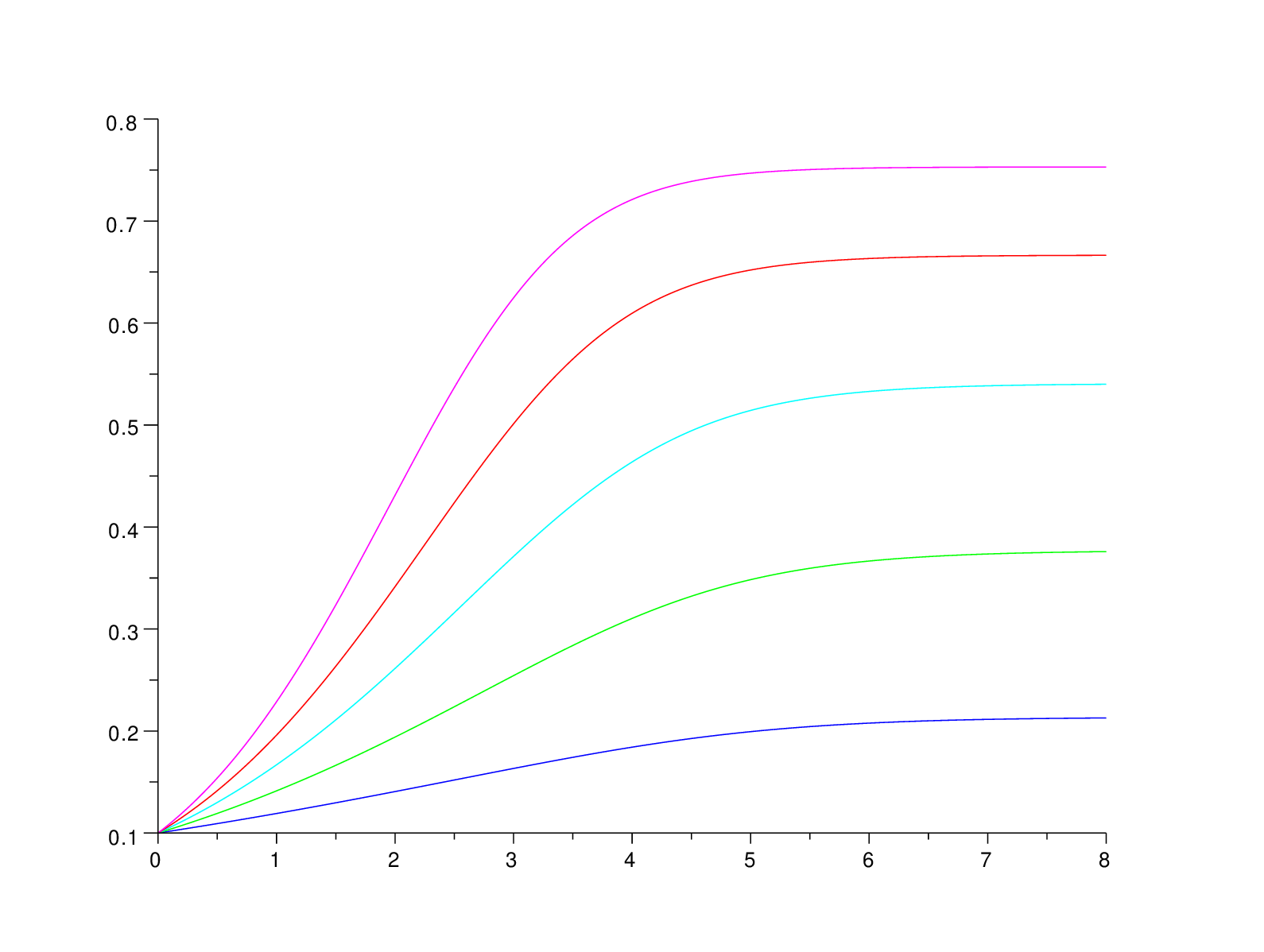}
} 
\caption{Evolution in time of the proportion of mutant cells
  in time in the Volterra model. The initial proportion of mutants is
  $0.1$. The IDR of normal cells is $1$. That of
  mutant cells is $1.2$ (blue), $1.4$ (green), $1.6$ (light blue),
  $1.8$ (red), $2$ (purple).}
\label{fig:VMpro}
\end{figure}

\begin{figure}[!ht]
\centerline{
\includegraphics[width=10cm]{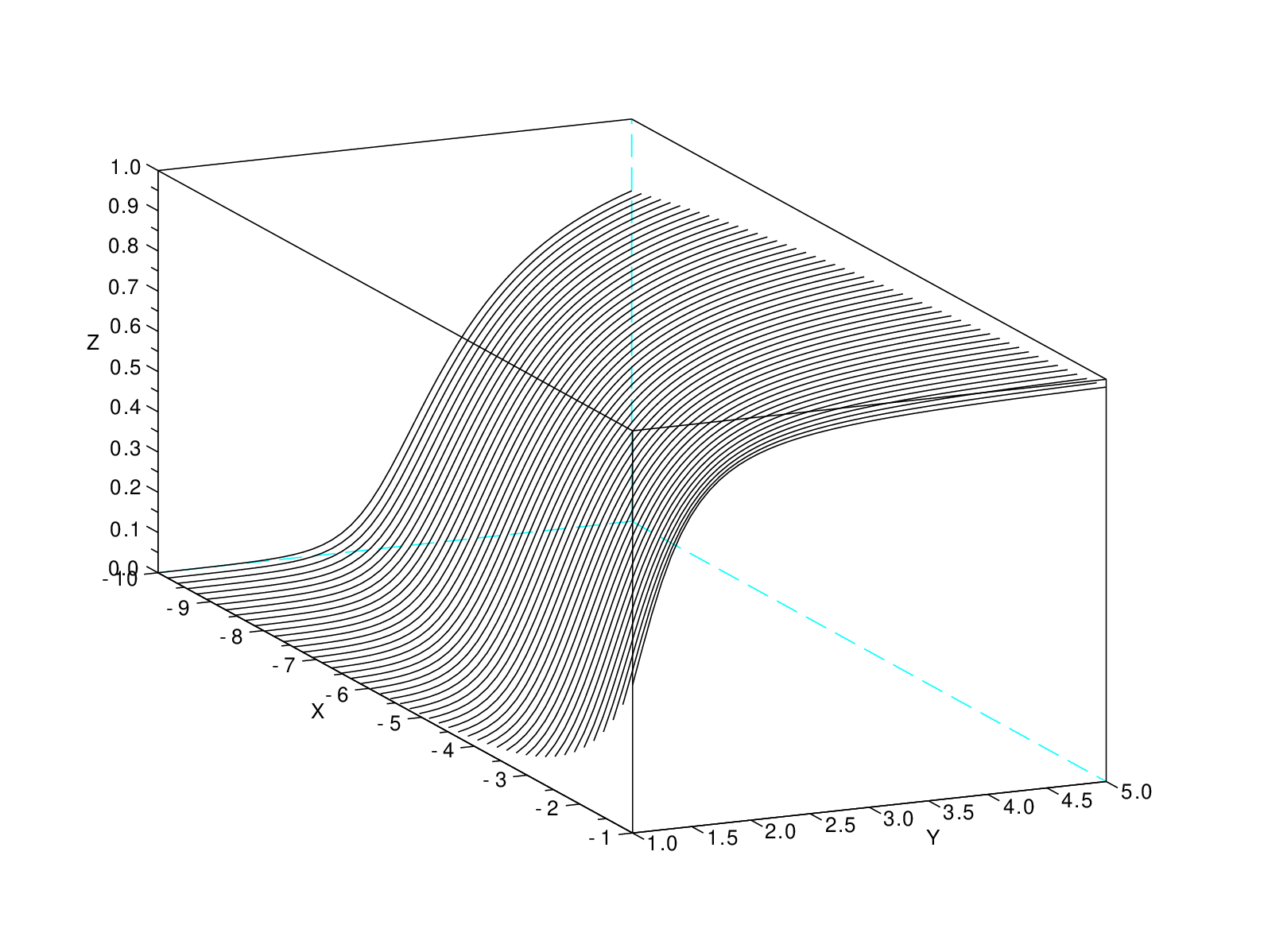}
} 
\caption{Limit proportion of mutant cells in the Volterra model
  (Z-axis, normal scale). 
The initial proportion $P(0)$, ranges from $\ee^{-1}=0.368$ to 
$\ee^{-10}=4.5\times 10^{-5}$
(X-axis, log-scale). The 50 fitnesses are linearly spaced between $1$
and $5$ (Y-axis, normal scale).}
\label{fig:VMlimpro}
\end{figure}

\begin{figure}[!ht]
\centerline{
\includegraphics[width=10cm]{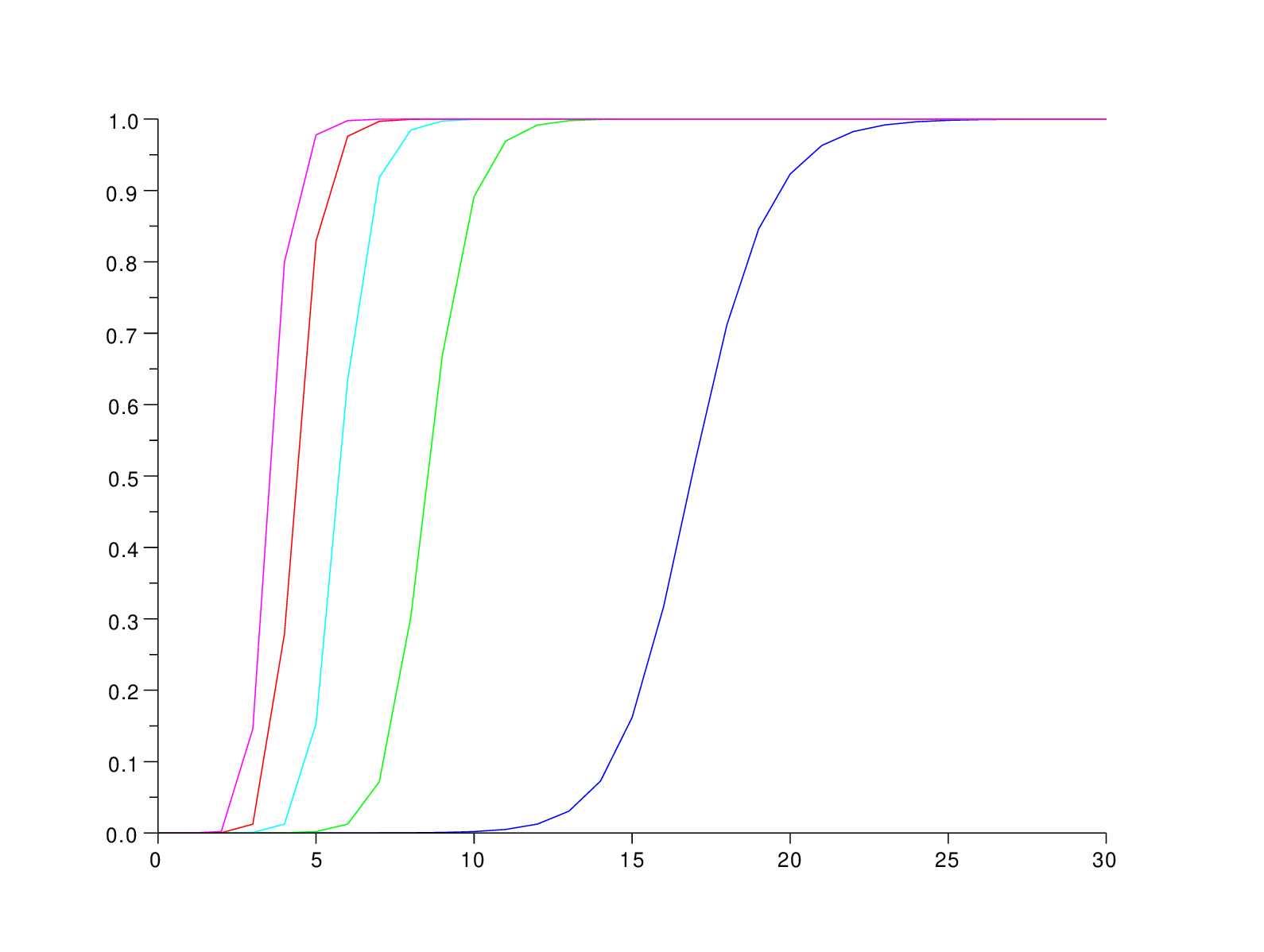}
} 
\caption{Daily evolution for the proportion of mutant cells. 
The initial proportion is $1/(5\,10^6)$ at the beginning of day $0$.
The IDR of normal cells is $1$. That of
  mutant cells is $1.2$ (blue), $1.4$ (green), $1.6$ (light blue),
  $1.8$ (red), $2$ (purple).}
\label{fig:VMday}
\end{figure}

The simplest deterministic model, that of exponential growth, is named
after Malthus \cite{Malthus1798}:
$$
\frac{\dd N(t)}{\dd t} = \nu N(t)\;,
$$
where $N(t)$ denotes the number of cells at time $t$, and 
$\nu$ is the individual division rate (IDR) of each cell.
That model had been considered long before Malthus, in particular by
Euler. What Malthus actually discussed in 1798 was precisely the
growth limitation due to lack of resources. 
\begin{quote}
It may safely be pronounced, therefore, that population, when
unchecked, goes on doubling itself every twenty-five years, or
increases in a geometrical ratio.

The rate according to which the productions of the earth may be
supposed to increase, it will not be easy to determine. Of this,
however, we may be perfectly certain, that the ratio of their increase
must be totally of a different nature from the ratio of the increase
of population.  
\end{quote}
Quetelet was the first one to propose to
account for growth limitations by substracting a quadratic 
correction term to the differential equation above, in 1836 (see
\cite{Quetelet1836} p.~288). 
\begin{quote}
La population tend \`a cro\^\i tre selon une progression
g\'eom\'etrique.

La r\'esistance, ou la somme des obstacles \`a son d\'eveloppement,
est, toutes choses \'egales d'ailleurs, comme le carr\'e de la vitesse
avec laquelle la population tend \`a cro\^\i tre.
\end{quote}
Two years later, Verhulst \cite{Verhulst1838} 
discussed various other limitating terms and compared the
models so obtained to experimental data. The model with a
quadratic limitation is now known as Verhulst model, or else
 \emph{logistic model}.
$$
\frac{\dd N(t)}{\dd t} = \nu N(t)\left(1-\frac{N(t)}{n_f}\right)\;,
$$
where $n_f$ is the maximal population sustained by the environment.
The model was generalized to several species competing in the same
environment by Lotka \cite{Lotka25} in 1925, and Volterra 
\cite{Volterra26} in
1926. Volterra discussed the different models in a course given in
1928 to the newly founded Institut Henri Poincar\'e. The lecture notes
``Lessons on the mathematical theory of the struggle for life''
\cite{Volterra31}, to which we shall refer, appeared in 1931. Soon
after, Lotka and Volterra predictions were confronted to
experimental data coming from all sorts of different contexts, in
particular by Gause \cite{Gause34}.
\vskip 2mm
Here are the first lines of Volterra \cite{Volterra31}, from
section 1.1 entitled ``Deux esp\`eces se disputant la m\^eme
nourriture''.
\begin{quote}
Supposons que, avec une nourriture en quantit\'e suffisante pour
satisfaire compl\`etement la voracit\'e de ces \^etres, il y ait des
coefficients d'accroisse\-ment positifs et constants $\varepsilon_1$,
$\varepsilon_2$. Si nous nous pla\c{c}ons maintenant dans le cas r\'eel
d'esp\`eces vivant dans un milieu d\'elimit\'e, la nourriture
diminuera quand les nombres $N_1$, $N_2$ des individus des deux
esp\`eces augmenteront et cela fera baisser la valeur des coefficients
d'accroissement. Si l'on repr\'esente la nourriture d\'evor\'ee par
unit\'e de temps par $F(N_1,N_2)$ fonction nulle avec $N_1$, $N_2$
ensemble, tendant vers l'infini avec chacune des variables et fonction
croissante de chacune d'elles, il sera assez naturel de prendre comme
coefficients d'accroissement
$$
\varepsilon_1-\gamma_1F(N_1,N_2)\;,\qquad
\varepsilon_2-\gamma_2F(N_1,N_2)\;,
$$
$\gamma_1$, $\gamma_2$ \'etant des constantes positives correspondant
aux deux esp\`eces et \`a leurs besoins respectifs de nourriture.

D'o\`u le syst\`eme diff\'erentiel traduisant le d\'eveloppement des
esp\`eces.
\begin{equation*}
\tag{1}
\frac{\dd N_1}{\dd t}=[\varepsilon_1-\gamma_1F(N_1,N_2)]N_1\;,\qquad
\frac{\dd N_2}{\dd t}=[\varepsilon_2-\gamma_2F(N_1,N_2)]N_1\;.
\end{equation*}
Maintenant se pose le probl\`eme math\'ematique d'\'etudier les
int\'egrales $N_1$, $N_2$ de ce syst\`eme, avec des valeurs initiales
$N_1^0$, $N_2^0$ positives pour $t=t_0$.

On peut d\'emontrer que pour tout intervalle fini $(t_0,T)$ il y a une
solution unique, de deux fonctions continues, restant comprises entre
deux nombres positifs, le plus grand ne d\'ependant pas de
l'extr\'emit\'e $T$ de l'intervalle (c'est-\`a-dire que $N_1$, $N_2$
restent born\'es).

\'Etudions ce qui arrive quand le temps s'\'ecoule ind\'efiniment. En
transcrivant (1) sous la forme 
\begin{equation*}
\tag{1'}
\frac{\dd \log N_1}{\dd t}=\varepsilon_1-\gamma_1F(N_1,N_2)\;,\qquad
\frac{\dd \log N_2}{\dd t}=\varepsilon_2-\gamma_2F(N_1,N_2)\;,
\end{equation*}
il vient par combinaison
$$
\gamma_2 \frac{\dd \log N_1}{\dd t}
-\gamma_1\frac{\dd \log N_2}{\dd
  t}=\varepsilon_1\gamma_2-\varepsilon_2\gamma_1\;,
$$
puis
\begin{equation*}
\tag{2}
\frac{N_1^{\gamma_2}}{N_2^{\gamma_1}}
=
\frac{(N_1^0)^{\gamma_2}}{(N_2^0)^{\gamma_1}}\,
\ee^{(\varepsilon_1\gamma_2-\varepsilon_2\gamma_1)(t-t_0)}\;.
\end{equation*}
N\'egligeons le cas infiniment peu probable o\`u 
$$
\varepsilon_1\gamma_2-\varepsilon_2\gamma_1=0
$$
et supposons, en permutant au besoin les esp\`eces, que
$$
\varepsilon_1\gamma_2-\varepsilon_2\gamma_1>0
\quad\mbox{ou}\quad
\frac{\varepsilon_1}{\gamma_1}>\frac{\varepsilon_2}{\gamma_2}\;;
$$
alors, d'apr\`es (2),
\begin{equation*}
\tag{3}
\lim_{t=\infty}
\frac{N_1^{\gamma_2}}{N_2^{\gamma_1}}
=+\infty\;.
\end{equation*}
$N_1$ restant born\'e, $N_2$ tend donc vers $0$.

Nous concluerons donc que \emph{la seconde esp\`ece, celle de
  $\frac{\varepsilon}{\gamma}$ le plus petit, s'\'epuise} et 
dispara\^\i t, \emph{tandis que la premi\`ere subsiste}.
\end{quote}
More as an example than a realistic model, we shall consider a
particular case of Volterra's system. Let us denote by $N(t)$ the
number of normal cells at time $t$, and by $M(t)$ the number of mutant
cells. Let $\nu$ and $\mu$ be their respective IDR's, 
and $n_f$ denote the maximal capacity (total number of cells
sustained by the medium). Our model, that we shall call \emph{Volterra
Model} (VM) even though it is only a very particular case of (1) deemed
``infiniment peu probable'' by Volterra, will be:
\begin{equation*}
\tag{VM}
\left\{
\begin{array}{lcl}
\displaystyle{\frac{\dd N(t)}{\dd t}}&=&
\displaystyle{\nu N(t)\left(1-\frac{N(t)+M(t)}{n_f}\right)}\\[2ex]
\displaystyle{\frac{\dd M(t)}{\dd t}}&=&
\displaystyle{\mu M(t)\left(1-\frac{N(t)+M(t)}{n_f}\right)}
\end{array}
\right.
\end{equation*}
In other words, the growth of both normal and mutant cells is equally
restricted, proportionally to their sum. It stops when that sum
reaches the maximal capacity $n_f$. 
Of course, the IDR's $\nu$ and $\mu$ should be estimated, and we
shall review estimation methods in section \ref{estimation}. For the
moment, we shall assume that the time scale has been set so that
$\nu=1$. Therefore $\mu=\mu/\nu$ can be viewed as the
\emph{fitness} of mutants, and we shall mainly consider here the case
$\mu\geqslant 1$. 
\vskip 2mm
No explicit solution can be given to the system of differential equations
(VM). However, it is quite easy to compute a numerical solution. Figure
\ref{fig:VMtraj} plots typical trajectories for $N(t)$ and $M(t)$,
all starting from the same initial values $N(0)=0.9\times(5\times 10^{6})$ and
$M(0)=0.1\times (5\times 10^6)$. The value of 
$\mu$ (the relative fitness of mutants) ranges from
$1.2$ to $2$. All curves have the same logistic-type shape. They
increase exponentially at first (when the nutrient limitation is not
yet significant), then converge asymptotically to a limit value as $t$
tends to $+\infty$. The calculations show that the
asymptotic value is reached in practice 
for relatively small values of $t$. All numerical calculations and
plots were made through Scilab \cite{Scilab}.
\vskip 2mm
Let us
denote by $N(\infty)$ and $M(\infty)$, the
limits as $t$ tends to infinity of $N(t)$ and $M(t)$. They can be computed
using Volterra's method:
$$
\left\{
\begin{array}{lcl}
\displaystyle{\frac{(N(\infty))^\mu}{(M(\infty))^\nu}}&=&
\displaystyle{\frac{(N(0))^\mu}{(M(0))^\nu}}\\[2ex]
\displaystyle{N(\infty)+M(\infty)}&=&n_f
\end{array}
\right.
$$
\vskip 2mm
Our main interest is in the evolution of the relative
\emph{proportion} of mutants at time $t$, denoted by $P(t)$:
$$
P(t) = \frac{M(t)}{N(t)+M(t)}\;.
$$
Figure \ref{fig:VMpro} shows the evolution in time of $P(t)$, with the
same values of the parameters as before.
Unlike the general case studied by Volterra, the two populations reach
an equilibrium where the proportion of mutants is not $1$. Nevertheless
the IDR of mutants being larger than that of normal cells, $P(t)$
increases in time. It converges to an asymptotic value, denoted by
$P(\infty)$. That ``limit proportion of mutants'' can be viewed 
as a function both of the initial proportion $P(0)$ and the 
fitness $\mu/\nu$. Figure \ref{fig:VMlimpro} plots that
function, for small values of $P(0)$ and fitnesses ranging from $1$ to $5$.
\vskip 2mm
Consider a given strain of mutants with fitness $\mu>1$, 
and denote by $f_\mu$ the (increasing) function that maps $P(0)$ onto
$P(\infty)$. Say that the mutation has taken place on day $0$, and
denote by $P_0$ the (small but positive) proportion 
of mutant cells at the end of day $0$. After dilution, the proportion
of mutant cells at the beginning of day $1$ will be a random variable
with expectation $P_0$ 
(see the discussion in section \ref{sampling}). Provided it is non
null, at the end of day $1$
the new proportion will be the image by $f_\mu$ of the initial
one. Dilution transforms it in
another random variable $P_1$ with expectation $f_\mu(P_0)$, and so
on. Denote by $P_n$ the proportion of mutants at the beginning of day
$n$. At the end of day $n$, it will be $f_\mu(P_n)$, then at the
beginning of day $n+1$, $P_{n+1}$ will be a new random variable with
expectation $f_\mu(P_n)$. The sequence of random variables so defined
is a Markov chain, converging to $1$ (mutants will eventually invade
the whole population). Using renewal theory techniques, the
distribution of the (random) number of days
before complete invasion can be precisely estimated. The rapid growth
of $f_\mu$ (figure \ref{fig:VMlimpro}) makes it quite likely that even
starting at day $0$ with a small proportion of mutants, the number of
days to complete invasion will be relatively small, conditionally of
course to the fact that mutants do not disappear upon dilution in the
first days. An estimate for the number of days until complete invasion
can be obtained by iteratively applying $f_u$, starting with $p_0=1/n_0$
(a single mutant on day $0$). The results are plotted on figure
\ref{fig:VMday}: with a relative fitness of 1.2, it takes about 25 days
for mutant cells to invade the population, whereas with a relative
fitness of 2, it takes 5 days.
\vskip 2mm
Symmetrically, the same techniques will show that for a
nonbeneficial mutation ($\mu\leqslant 1$), 
there are very little chances that the
mutation will survive day $2$, even if it survives dilution on day $1$.
\vskip 2mm
A more sophisticated model, adapted to the context of
bacterial growth experiments, was introduced by 
Herbert, Ellsworth \& Telling, and independenty by Powell in 1956 (see
\cite{Powell58,Hsu77}, and references therein). A description is
given in chapter 12 of Kot \cite{Kot01}. The model takes into account
the nutrient, and follows the evolution of bacterial cells as a
Michaelis-Menten enzymatic reaction (see \cite{JohnsonGoody11} for a
historical account of Michaelis \& Menten's original paper). To the
number of normal and mutant cells $N(t)$ and $M(t)$, one adds the
substrate (nutrient) 
concentration $S(t)$. The model, that we shall call 
(somewhat arbitrarily) \emph{Powell Model} or PM is given by the following
system of differential equations
(see Kot \cite{Kot01}, system (12.16) p.~205). 
\begin{equation*}
\tag{PM}
\left\{
\begin{array}{lcl}
\displaystyle{\frac{\dd S(t)}{\dd t}}&=&
\displaystyle{D(S_i-S(t))-\frac{1}{y_n}\frac{\nu S(t)N(t)}{k_n+S(t)}
-\frac{1}{y_m}\frac{\mu S(t)N(t)}{k_m+S(t)}}\\[2ex]
\displaystyle{\frac{\dd N(t)}{\dd t}}&=&
\displaystyle{\frac{\nu S(t)N(t)}{k_n+S(t)}-D N(t)}\\[2ex]
\displaystyle{\frac{\dd M(t)}{\dd t}}&=&
\displaystyle{\frac{\mu S(t)M(t)}{k_m+S(t)}-D N(t)}
\end{array}
\right.
\end{equation*}
The parameters are the following.
\begin{itemize}
\item $\nu,\mu$: the IDR's of normal and mutant cells,
\item $y_n,y_m$: the yield coefficients of normal and mutant cells
  (consumption of substrate by bacteria per unit of time),
\item $k_n,k_m$: the half-saturation constants of normal and mutant
  cells,
\item $D$: the dilution,
\item $S_i$: the inflowing substrate concentration.
\end{itemize}
In our case, one can consider that $S_i=0$ for each single day. Besides being
closer to the experimental reality, the PM has another interesting
feature. In the VM we considered that the evolutionary advantage of the
mutants translated only into a faster division rate. It might also
yield a higher consumption of nutrients, which would even
accelerate the invasion. A mathematical study of the PM is no more
difficult than for the VM; but it would require estimates for
7 parameters instead of 3.
Nevertheless, we believe that the qualitative
conclusions that were drawn from the VM would still remain true for
the PM; they are summarized below.
\begin{enumerate}
\item
nonbeneficial mutations disappear over a few days, through the effect
of daily dilution
\item
beneficial mutations, if the strain survives dilution on the first
day, have a high
probability to invade the full population, and eventually to eliminate
normal cells, after at most a few tens of days.
\item
On a single day, two competing strains of bacteria increase at
different rates, according to their relative fitnesses; their
populations eventually stabilize to asymptotic values, the relative
proportion of the fittest being strictly larger at the end of the day 
than at the beginning.   
\end{enumerate}
We also wish to point out that the models above can be easily extended
to more than 2 competing species. As a common feature, the generalizations
will obey the \emph{principle of competitive exclusion} (see
section 3.5 of Murray \cite{Murray89}): in the long term, the fittest
species eliminates the others. 
\vskip 2mm
The fixation along successive generations of beneficial mutations has
been discussed in many references and textbooks. Papp \emph{et al.}
\cite{Pappetal11} recently wrote an interesting review on 
systems-biology approaches to genomic evolution.
Lang \emph{et al.}
\cite{Langetal11} propose a thorough experimental study in different
strains of yeast. Campos \& Wahl \cite{CamposWahl09,CamposWahl10}
discuss the effect of population bottlenecks. Hermisson \&
Pfaffelhuber \cite{Hermisson07} study genetic hitchhiking under
recurrent mutations.
\section{Stochastic growth models}
\label{birth}
\begin{figure}[!ht]
\centerline{
\includegraphics[width=10cm]{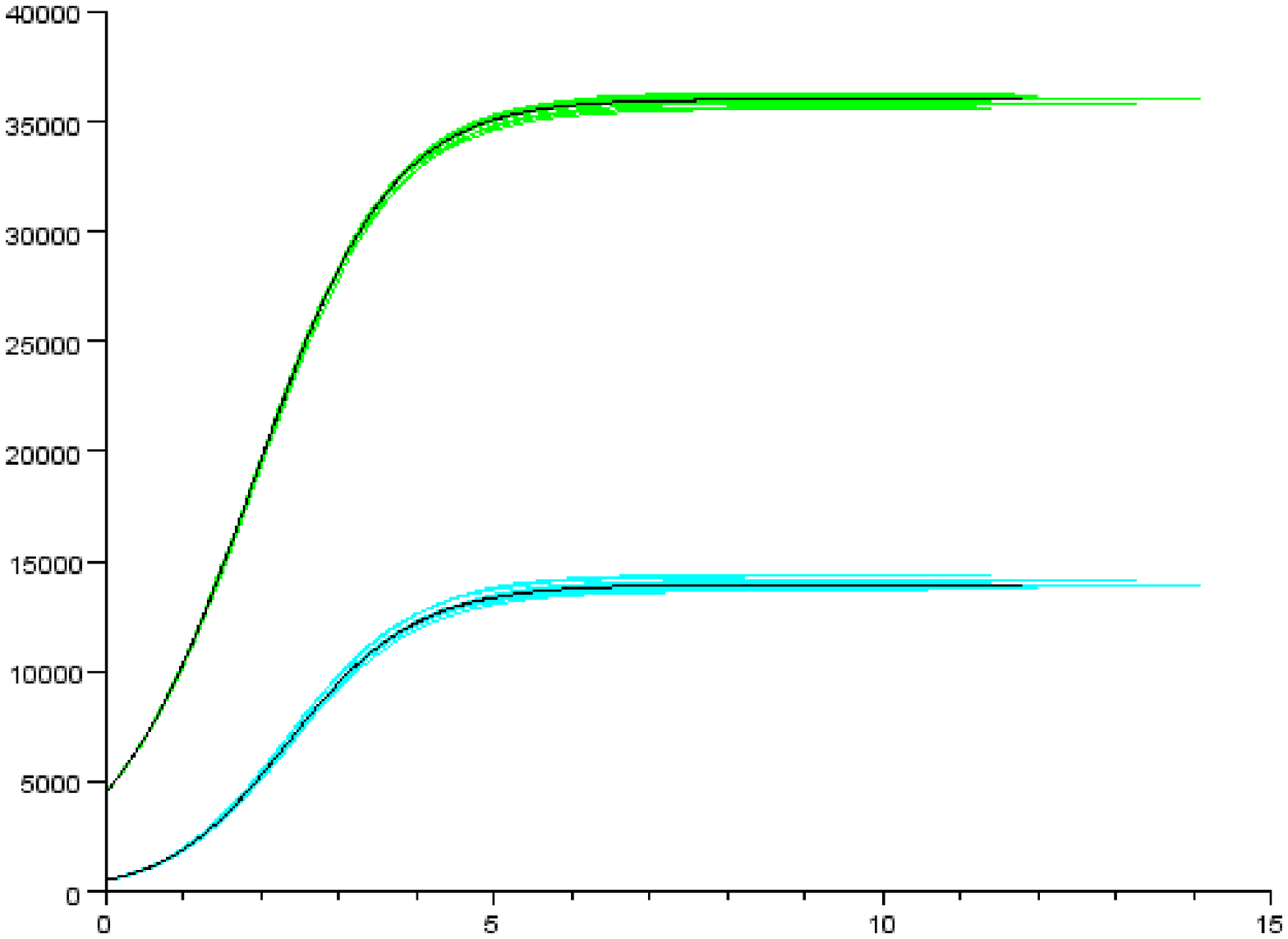}
} 
\caption{Ten trajectories of the stochastic competition model
for the numbers of normal (green) and mutant cells (blue). The
trajectory of the corresponding deterministic model is plotted in
black. The initial and final numbers of cells are $n_0=5\times 10^3$ and
$n_f=5\times 10^4$. The initial proportion of mutant cells is $0.1$.
The IDR's of normal and mutant cells are $1$ and $1.6$.}
\label{fig:CMtraj}
\end{figure}

The probabilistic modelling of
population growth started in 1925 with Yule \cite{Yule25}. Such an
early date is somewhat misleading. The burst of the theory
took place in a few years around 1950 with such important
contributors as Bartlett \cite{Bartlett49,Bartlett51,Bartlett54},
Harris \cite{Harris51,BellmanHarris52} and Kendall
\cite{Kendall48,Kendall49,Kendall52}. More references and a
perspective on the early historical development can be found in the
discussion following Armitage's presentation to the Royal Statistical
Society \cite{Armitage52}, the chairman being
Bartlett. Kendall \cite{Kendall52} gives another useful early review. Of
course the same material has since appeared in many textbooks, starting
with Bartlett's \cite{Bartlett60,Bartlett66} (see also Bharucha-Reid
\cite{Bharucha60}). Since the 1950's, stochastic population models have
remained a lively subject: see \emph{e.g.} Pakes \cite{Pakes03} for a
more recent review. Aldous \cite{Aldous01} gives an interesting
perpective relating Yule's founding paper to present researches.
\vskip 2mm
The basic hypothesis of all models is that all bacteria behave
independently in the same manner; of course this should be understood in
the stochastic sense: the times at which all bacteria divide are
independent and identically distributed random variables. The common
distribution function of these random times will be denoted by $F$. The
two simplest particular cases are:
\begin{itemize}
\item \emph{deterministic:} $F(t)=\II_{[t_0,+\infty)}(t)$ (the
  division time is fixed and equal to $t_0$).
\item \emph{Markovian:} $F(t)=(1-\ee^{-\nu t})\II_{[0,+\infty)}(t)$
(the division time is exponentially distributed with parameter
$\nu$).
\end{itemize}
These are the cases where an explicit mathematical treatment is easily
feasible, and we shall focus on the second one. More general families of
distributions have been considered, including Gamma distributions (see
Kendall \cite{Kendall48,Kendall52} for a discussion and comparison
with experimental data).
\vskip 2mm 
 When a division occurs, the cell that divides 
is turned into a random number of
new bacteria, each having a fresh division date. Let $\varphi$ be the
generating function of the random number of ``children'' produced at
division time:
$$
\varphi(z)=\EE[z^K]=\sum_{k=0}^{+\infty} z^k\,\PP[K=k]\;,
$$
where $K$ is the random number of children at any division. 
Even though we shall be concerned exclusively by the deterministic
case $K\equiv 2$ (or $\varphi=z^2$), let us mention that the general
case is the basis of the famous Galton-Watson model of branching
processes. Notice also that $K$ can be null, which corresponds to a death.
\vskip 2mm
Generating functions of
integer valued random variables will
be used extensively in what follows. Let us recall the following
basic properties:
\begin{enumerate}
\item 
If two random variables $N$ and $M$ are independent, then 
the generating function of
  $N+M$ is the product of the generating functions of $N$ and $M$.
\item
The generating function of $N$ has a $k$-th left derivative at $z=1$
if and only if $\EE[N^k]$ exists. In that case,
$$
\lim_{z\to1^-}\frac{\dd \EE[z^N]}{\dd z}=\EE[N(N-1)\cdots(N-k+1)]\;.
$$ 
\end{enumerate}
The basic object of interest is the number of bacteria alive at time
$t$, when one single cell is present at time $0$; 
its generating function will be denoted by
$G(z,t)$.
$$
G(z,t)=\sum_{n=0}^{+\infty} z^n\,\PP[N_t=n\,|\,N_0=1]\;,
$$
where $N_t$ denotes the (random) number of bacteria living at time
$t$, under the hypothesis that $N_0\equiv 1$. 
The function $G$ is related to $F$ and $\varphi$ through the
\emph{Bellman-Harris integral equation} (see \cite{BellmanHarris52}). 
\begin{equation*}
\tag{BH}
G(z,t) = \left(\int_0^t \varphi(G(z,t-s))\,\dd F(s)\,\right)+z(1-F(t))\;.
\end{equation*}
Even though we shall not provide a rigorous proof, giving an
intuitive interpretation of that equation can still be useful for future
developments. Let us argue as follows. The distribution function $F(t)$ is the
probability that the division of the initial cell has taken
place no later than $t$; $1-F(t)$ is the probability it has occurred
after $t$, $\dd F(s)$ is the probability that it occurs between $s$
and $s+\dd s$. If at time $t$ the division has not occurred
(probability $(1-F(t))$) then the number of bacteria is still equal to
$1$, and its generating function is $z$: this accounts for the second
part of the right hand side, $z(1-F(t))$. Suppose now that a division has
occurred at some time $s$ between $0$ and $t$ (in $[s,s+\dd s]$ with
probability $\dd F(s)$). Assume it
leads to $k$ new cells, each starting a new lineage. Since all
lineages develop with identical rules, the population
at time $t$, stemming from one lineage which started at time $s$, will have
generating function $G(z,t-s)$. The total number will be the sum of
all $k$ lineages, which are supposed to be independent: its generating
function will be the $k$-th power $G^k(z,t-s)$. Now since $k$
cells are produced with probability $\PP[K=k]$, the generating
function at time $t$ becomes
$$
\sum_{k=0}^{+\infty} G^k(z,t-s)\,\PP[K=k]=\varphi(G(z,t-s))\;.
$$
This accounts for the first term in the (BH) equation.
\vskip 2mm
As a particular case, when 
$F(t)=(1-\ee^{\nu  t})\II_{[0,+\infty)}(t)$ 
(exponential distribution, Markovian case), and $K\equiv 2$
($\varphi(z)=z^2$), the Bellman-Harris equation (BH) becomes:
$$
G(z,t) = \left(\int_0^t G^2(z,t-s)\nu \ee^{-\nu s}\,\dd
  s\,\right)+
z\ee^{-\nu t}\;.
$$
Inside the integral, change $t-s$ into $u$:
$$
G(z,t) = \left(\,\ee^{-\nu t}
\int_0^t G^2(z,u)\nu \ee^{\nu u}\,\dd
  u\,\right)+
z\ee^{-\nu t}\;.
$$
or else:
$$
G(z,t) \ee^{\nu t}=\left(\,
\int_0^t G^2(z,u)\nu \ee^{\nu u}\,\dd
  u\,\right)+
z\;.
$$
The derivative in time is:
$$
\frac{\partial G(z,t)}{\partial t} \ee^{\nu t} +
G(z,t)(\nu\ee^{\nu t}) = 
G^2(z,t)(\nu \ee^{\nu t})\;.
$$
Simplifying by $\ee^{\nu t}$ and rearranging terms, leads to the
following differential equation.
\begin{equation*}
\tag{DE}
\frac{\partial G(z,t)}{\partial t} = \nu(G^2(z,t)-G(z,t))\;.
\end{equation*}
The solution to (DE) for $G(z,0)=z$ is easily computed:
$$
G(z,t)= \frac{\ee^{-\nu t} z}{1-(1-\ee^{-\nu t})z}\;.
$$
This is the generating function of the 
geometric distribution with parameter $p=\ee^{-\nu  t}$. 
In other terms, for all $n\geqslant 1$, the probability that
$n$ bacteria are alive at time $t$ is:
$$
\PP[N_t=n] = \ee^{-\nu t}(1-\ee^{-\nu t})^{n-1}\;.
$$
The first two moments of $N_t$ are:
$$
\EE[N_t] = \frac{1}{p} = \ee^{\nu t}
\qquad\mbox{and}\qquad
\mbox{Var}[N_t] = \frac{1-p}{p^2} = \ee^{2\nu t}-\ee^{\nu t}\;.
$$
The Markov process $\{N_t\,,\;t\geqslant 0\}$ is one of the simplest
examples of a birth-and-death process (actually a ``pure birth'' process).
It is usually named after Yule (from \cite{Yule25}), 
who derived the geometric distribution of
$N_t$. Furry \cite{Furry37} introduced the same process in another
context.
Novozhilov \emph{et al.} \cite{Novozhilov05} give a
very simple presentation of birth-and-death processes used in 
biology. 
\vskip 2mm
In the LE, the number of bacteria varies from $n_0=5\times 10^6$ to
$n_f=5\times 10^8$. So it is legitimate to consider 
large number approximations. 
If $n_0$ bacteria are initially present, their lineages
evolve independently, according to the distribution that has just been
described: at each time $t$, the total population is distributed as
the sum of $n_0$ independent random variables, each following the
geometric distribution with parameter $\ee^{-\nu t}$, that is a
Pascal distribution with parameters $n_0$ and $\ee^{-\nu t}$. If
$n_0$ is large, it can be
approximated by a Gaussian (normal) distribution with same mean and
variance, \emph{i.e.}:
$$
\EE[N_t] = \frac{n_0}{p} = n_0\ee^{\nu t}
\qquad\mbox{and}\qquad
\mbox{Var}[N_t] = n_0\frac{1-p}{p^2} = n_0\ee^{2\nu t}-n_0\ee^{\nu t}\;.
$$
Observe that for $n_0$ large, the standard-deviation is small compared
to the expectation:
$$
\EE[N_t] = \frac{n_0}{p} = n_0\ee^{\nu t}
\qquad\mbox{and}\qquad
\sqrt{\mbox{Var}[N_t]} \simeq \sqrt{n_0\ee^{2\nu t}}=
\frac{\EE[N_t]}{\sqrt{n_0}}\;.
$$
Therefore, as a first approximation, the population is expected to
grow deterministically, as $n_0\ee^{\nu t}$.
\vskip 2mm
The Yule process is the stochastic counterpart of the deterministic 
exponential growth model. Let us denote by $N(t)$ and $n(t)$ the
respective numbers of cells
in the stochastic model and in the deterministic one.
The deterministic model is specified by an ordinary differential
equation.
$$
\frac{\dd n(t)}{\dd t} = \nu\,n(t)\;,
$$
or equivalently by the integral equation
$$
n(t) = n_0 + \int_0^t \nu\,n(s)\,\dd s\;.
$$
The basis of the stochastic model is a random time scale, specified by
a Poisson process. Let $\{Y(t)\,,\;t\geqslant 0\}$ be a unit Poisson
process: $Y$ has jumps of size $1$ at successive instants,
separated by exponentially distributed durations with expectation $1$.
The stochastic process $N(t)$ can be defined by the following integral
equation, which is the stochastic counterpart of the deterministic
one.
$$
N(t) = n_0+Y\left(\int_0^t \nu N(s)\,\dd s\right)\;.
$$
The distribution at time $t$ of $N(t)$ is the solution of a system of
ordinary differential equations, the \emph{Chapmann-Kolmogorov} equations
(also called the \emph{Chemical Master Equation} in the context of
kinetics). Denoting by $p_n(t)$ the probability
that $N(t)=n$, the equation for $n>0$ is:
$$
\frac{\dd p_n(t)}{\dd t} = -\nu n p_n(t)+\nu(n-1) p_{n-1}(t)\;.
$$
As we have seen, that system has an explicit solution. However, it is
a very particular case and no explicit solution can be hoped for in
more general situations. Also, the fact that the expectation of $N(t)$
is exactly equal to the deterministic solution $n(t)$ is quite rare. What is
general however, is the large number approximation that was
explained above. We shall rewrite it in a way that can be easily 
generalized. Its foundation is the long term approximation for the
Poisson process $Y$. At (large) time $n_0 u$, $Y(n_0u)$ equals $n_0 u$ on
average, with fluctuations of order $\sqrt{n_0}$, described by a
Brownian motion.
$$
\lim_{n\to +\infty} \frac{Y(n_0u)-n_0u}{\sqrt{n_0}} = W(u)\;,
$$
where $Y$ is a unit Poisson process, $W$ is the standard Brownian
motion and the limit is understood in distribution.
Consider now the integral equation defining $N(t)$ and divide both
members by $n_0$. 
$$
\frac{N(t)}{n_0} = 1+\frac{1}{n_0}Y\left(\int_0^t \nu N(s)\,\dd s\right)\;.
$$
Denote by $U(t)$ the ratio $\frac{N(t)}{n_0}$.
$$
U(t) = 1+ \frac{1}{n_0}Y\left(\int_0^t \nu n_0U(s)\,\dd s\right)\;.
$$
Approximating $Y(n_0u)$ by $n_0 u+\sqrt{n_0}W(u)$:
$$
U(t) \simeq 1+ \int_0^t \nu U(s)\,\dd s
+\frac{1}{\sqrt{n_0}}W\left(\int_0^s \nu U(s)\,\dd s\right)\;.
$$
The diffusion process $U$ is the solution of a \emph{Stochastic
  Differential Equation} (called the \emph{Chemical Langevin Equation}
in kinetics)
$$
\dd U(t) = \nu U(t)\,\dd t+\frac{1}{\sqrt{n_0}} \sqrt{\nu U(t)}\,\dd W(t)\;.
$$
Of course for very large $n_0$, the diffusion term vanishes and the
equation becomes the deterministic differential equation
(the \emph{Reaction Rate Equation} in kinetics).
\vskip 2mm
What we have just seen on the example of the Yule process is an
illustration of a very general modelling principle. Three different 
\emph{scales} can be considered.
\begin{enumerate}
\item \emph{Microcopic scale}: stochastic jump process with
  discrete state space. The random fluctuations are governed by
  Poisson processes;
\item \emph{Mesoscopic scale}: stochastic diffusion process with
  continous state space. The random fluctuations are governed by
  Brownian motions;
\item \emph{Macroscopic scale}: deterministic function of time,
  solution of a differential system of equations.
\end{enumerate}
The three scales are coherent in the sense that each scale is a large
number approximation of the previous one. 
\vskip 2mm
In order to illustrate the three scales principle, we shall introduce
here a microscopic model of competition between normal and mutant
cells, matching the Volterra model studied in section \ref{volterra}.
Let $N(t)$ and $M(t)$ still denote the numbers of normal and mutant
cells. At each division, one of the two counts increases by one unit. When
$(N(t),M(t))=(n,m)$, the respective rates of
increase of normal and mutant cells are: 
$$
\rho(n,m)=\nu n\left(1-\frac{n+m}{n_f}\right)
\quad\mbox{and}\quad
\sigma(n,m)=\mu m\left(1-\frac{n+m}{n_f}\right)\;.
$$
We shall call this model \emph{competitive process} (CP).
A loose description of the CP can be given as follows.
\begin{enumerate}
\item
When $n$ normal and $m$ mutant cells are present, the next division
will occur after a random time, following the exponential distribution
with parameter $\rho(n,m)+\sigma(n,m)$.
\item
Upon next division,
\begin{enumerate}
\item 
with probability $\frac{\rho(n,m)}{\rho(n,m)+\sigma(n,m)}$, $N(t)$
will increase by $1$, $M(t)$ remaining unchanged.
\item 
with probability $\frac{\sigma(n,m)}{\rho(n,m)+\sigma(n,m)}$, $M(t)$
will increase by $1$, $N(t)$ remaining unchanged.
\end{enumerate}
\end{enumerate}
The formal description uses two independent unit Poisson processes,
$Y_1$ and $Y_2$.
\begin{equation*}
\tag{CP}
\left\{\begin{array}{lcl}
N(t)&=&\displaystyle{N(0)+Y_1\left(\int_0^t 
\rho(N(s),M(s))\,\dd s\right)}\\[2ex]
M(t)&=&\displaystyle{M(0)+Y_2\left(\int_0^t 
\sigma(N(s),M(s))\,\dd s\right)}
\end{array}\right.
\end{equation*}
Let $p_0$ be the initial proportion of mutant molecules, so that
$N(0)=n_0(1-p_0)$ and $M(0)=n_0p_0$. Assuming that $n_0$ is large, we
reproduce the diffusion approximation scheme that was exposed above for
the Yule process. Dividing both equations by $n_0$,
then setting 
$$
U(t) = \frac{N(t)}{n_0}
\quad\mbox{and}\quad
V(t) = \frac{M(t)}{n_0}\;,
$$ 
one gets:
$$
\left\{\begin{array}{lcl}
U(t)&=&\displaystyle{1-p_0+\frac{1}{n_0}Y_1
\left(\int_0^t\nu n_0 U(s)
\left(1-\frac{n_0}{n_f}(U(s)+V(s))\right)\,\dd s\right)}\\[2ex]
V(t)&=&\displaystyle{p_0+\frac{1}{n_0}Y_2\left(\int_0^t 
\mu n_0 V(s)\left(1-\frac{n_0}{n_f}(U(s)+V(s)\right)\,\dd s\right)}
\end{array}\right.
$$
Replace now $Y_1(n_0u)$ and $Y_2(n_0v)$ by:
$$
n_0u+\sqrt{n_0}W_1(u)
\quad\mbox{and}\quad
n_0v+\sqrt{n_0}W_2(v)\;,
$$ 
where $W_1$ and $W_2$ are two independent standard Brownian motions.
The CM becomes:
$$
\left\{\begin{array}{lcl}
U(t)&\simeq&\displaystyle{1-p_0+\int_0^t 
\nu U(s)\left(1-\frac{n_0}{n_f}(U(s)+V(s)\right)\,\dd s}\\[1.5ex]
&&\displaystyle{
+\frac{1}{\sqrt{n_0}}W_1\left(\int_0^t
    U(s)\left(1-\frac{n_0}{n_f}(U(s)+V(s)\right)\,\dd s\right)}\\[2.5ex]
V(t)&\simeq&\displaystyle{p_0+\int_0^t 
\mu V(s)\left(1-\frac{n_0}{n_f}(U(s)+V(s)\right)\,\dd s}\\[1.5ex]
&&\displaystyle{
+\frac{1}{\sqrt{n_0}}W_2\left(\int_0^t
    V(s)\left(1-\frac{n_0}{n_f}(U(s)+V(s)\right)\,\dd s\right)}
\end{array}\right.
$$
Thus the stochastic process $(U,V)$ is a bidimensional diffusion,
solution to the following stochastic differential equations.
$$
\left\{\begin{array}{lcl}
\dd U(t)&=&\displaystyle{
\nu U(t)\left(1-\frac{n_0}{n_f}(U(t)+V(t)\right)\,\dd t}\\[1.5ex]
&&\displaystyle{+\frac{1}{\sqrt{n_0}}
    \sqrt{U(t)\left(1-\frac{n_0}{n_f}(U(t)+V(t))\right)}\dd W_1(t)}\\[2.5ex]
\dd V(t)&=&\displaystyle{ 
\mu V(t)\left(1-\frac{n_0}{n_f}(U(t)+V(t)\right)\,\dd t}\\[1.5ex]
&&\displaystyle{+\frac{1}{\sqrt{n_0}}
    \sqrt{\mu V(t)\left(1-\frac{n_0}{n_f}(U(t)+V(t))\right)}\dd W_2(t)}\\
\end{array}\right.
$$
This is the mesoscopic scale model. Of course, by neglecting the
diffusion term in $\frac{1}{\sqrt{n_0}}$, one gets the deterministic
(or macroscopic) Volterra model of section \ref{volterra}.
\vskip 2mm
How should one choose among the three modelling scales? Figure
\ref{fig:CMtraj} shows a simulation of 10 trajectories of the CM, for
$n_0=5\times 10^3$ and $n_f=5\times 10^4$, the relative fitness of
mutants being $1.6$.  
On the same graphics, the trajectory of the deterministic Volterra
model has been superposed. As can be observed, the dispersion of
random trajectories around the deterministic ones is small. For the
more realistic values of $n_0=5\times 10^6$ and
$n_f=5\times 10^8$, the random trajectories would be almost undistinguishable
from the deterministic ones. However, this is only true for a relatively
large value of $p_0$. If only a few mutant cells were initially
present, stochastic models would certainly give more reliable
results than the deterministic one.
\vskip 2mm
Once again, our intention in proposing that very simple model, was to
illustrate the possible treatments rather than imposing it as the most
realistic. More sophisticated logistic-type
stochastic models were proposed long ago: see
Novozhilov \cite{Novozhilov05} and section 6.8
p.~242 of Allen \cite{Allen03}. Logistic growth processes have been
the object of several mathematical studies, in particular by Tan \&
Piantadosi \cite{TanPiantadosi91}, or more recently by Lambert
\cite{Lambert05}.  Refinements include for instance spatial random
dispersal \cite{TuckwellKoziol87} or local regulation \cite{Etheridge04}.
\section{Mutation models}
\label{mutation}
The history of mutation models is as long as that of
stochastic growth processes, since estimates of mutation probabilities
were already present in Yule's founding paper \cite{Yule25}. However,
it really started with Luria and Delbr\"uck experiments
\cite{LuriaDelbruck43}. Interestingly enough, they used the same
\emph{Escherichia Coli B} as the LE. Their idea was to
introduce a virus that killed most bacteria, except those
having acquired resistance by mutation. This allowed them to count
mutant bacteria. Repeating the experiment, they estimated
the distribution of the (random) number of mutants. 
The data seemed to indicate a distribution with a much heavier
tail than expected: most counts had very few mutant cells, but in a
sizeable proportion of the counts, they found a relative high
number of mutants. In view of our discussion in section
\ref{sampling}, that feature is crucial for the survival of mutations
through daily sampling in the LE. Indeed we have shown that 
mutations have good chances to survive daily dilution, 
only when they are carried by
sufficiently many cells. Luria and Delbr\"uck used a very simple
deterministic model, assuming that cell counts doubled at each
multiple of a given fixed period. Nevertheless, that model allowed
them to derive an asymptotic distribution of the number of mutants,
that showed indeed a heavy tail behavior. Other models were later
consider, in particular by Lea \& Coulson \cite{LeaCoulson49}, Harris
\cite{Harris51}, Bartlett \cite{Bartlett49,Bartlett51}, Kendall
\cite{Kendall49,Kendall52}, and Armitage \cite{Armitage52}. Mandelbrot
\cite{Mandelbrot74} proposed a general proof for the convergence to the
Luria-Delbr\"uck distribution. More recently, the importance of that 
distribution for the treatment of mutation experimental data has
stemmed further researches such as
\cite{Sarkar91,Maetal92,Kemp94}. The Luria-Delbr\"uck distribution is
known only through its generating function and no close form exists
for its probabilities. However an efficient recursive formula permits
to calculate them explicitly. Pakes \cite{Pakes93} gives an easy
derivation of that formula.
An example of generalization is given by Dewanji \emph{et al.}
\cite{Dewanjietal05}. Presently, the most active author on the
subject is Zheng; he wrote a useful mathematical
review in \cite{Zheng99}. Historical reviews include Sarkar
\cite{Sarkar91} and Zheng \cite{zheng10}.
\vskip 2mm
The objective of this section is to present Bartlett's derivation of the
Luria-Delbr\"uck distribution under the following modelling
assumptions (see \cite{Bartlett51}, \cite{Armitage52}, p.~37, 
\cite{Bartlett66}, section 4.31 p.~124, and more recently Zheng 
\cite{Zheng08}).
\begin{enumerate}
\item At time $0$ a homogeneous population of $n_0$ ``normal'' cells is given
  ($n_0$ is large);
\item normal cells divide at a constant rate $\nu$;
\item when a division occurs:
\begin{enumerate}
\item with probability $1-p$ two normal cells are produced,
\item with probability $p$ one normal and one mutant cells are produced,
\end{enumerate}
(the mutation probability $p$ is small);
\item mutant cells divide at a constant rate $\mu$;
\item any other mutation than normal to mutant is excluded; 
\item all random events (divisions and mutations) are mutually independent.
\end{enumerate}
\vskip 2mm
Let $G$ denote the bivariate generating function for the numbers of
normal and mutant cells, starting with a single normal cell at
time $0$.
$$
G(z,y,t)=\sum_{n=0}^{+\infty}\sum_{m=0}^{+\infty} z^ny^m\,
\PP[N(t)=n\,,\;M(t)=m\,|\,N(0)=1\,,\;M(0)=0]\;.
$$
If the population starts with a single mutant cell, only mutant cells
can be produced later, and there is no need to consider a bivariate
generating function. We shall denote by $H$ the generating fonction of
the number of mutant cells, starting with a single mutant cell at $t=0$.
$$
H(y,t)=\sum_{m=0}^{+\infty} y^m\,\PP[M(t)=m\,|\,M(0)=1]\;.
$$
Its calculation from the Bellman-Harris equation was already 
exposed in section \ref{birth}.
$$
H(y,t)= \frac{y\ee^{-\mu t}}{1-y+y\ee^{-\mu t}}\;.
$$
The generating functions $G$ and $H$ are related
through another Bellman-Harris equation.
\begin{equation*}
\tag{BH2}
\begin{array}{lcl}
G(z,y,t) &=&\displaystyle{ \left(\int_0^t 
\Big((1-p)G^2(z,y,t-s)\right.}\\[2ex]
&&\displaystyle{\left.+p\,G(z,y,t-s)H(y,t-s)\Big)\nu \ee^{-\nu s}\,\dd
  s\,\right)+
z\ee^{-\nu t}\;.}
\end{array}
\end{equation*}
The justification of (BH2) is quite similar to that of (BH), given in
section \ref{birth}. The initial normal cell divides at a time which is
exponentially distributed with parameter $\nu$. At time $t$, it may not
have divided yet (with probability $\ee^{-\nu t}$), and the generating function
is still $z$. If it divides at some time
$s$ between $0$ and $t$ (in $[s,s+\dd s]$ with 
probability $\nu\ee^{-\nu s}\dd s$), it
turns either into 2 normal cells with probability $1-p$, or into a
normal and a mutant cell with probability $p$. The two new cells
start independent lineages of their own, accounted for by $G^2(z,y,t-s)$
if two normal cells are produced, and by $G(z,y,t-s)H(y,t-s)$ if one normal
and one mutant are produced. As in section \ref{birth}, 
(BH2) can be transformed into an ordinary differential
equation. 
\begin{equation*}
\tag{DE2}
\frac{\partial G(z,y,t)}{\partial t}= 
\nu((1-p)G^2(z,y,t)+p\,G(z,y,t)H(y,t)-G(z,y,t))\;.
\end{equation*}
This is a linear first order equation in the inverse $G^{-1}(z,y,t)$:
$$
\frac{\partial G^{-1}(z,y,t)}{\partial t}= 
\nu(p-1)+\nu G^{-1}(z,y,t)(1-p\,H(y,t))\;.
$$
Replacing $H(y,t)$ by its explicit expression, 
the general solution to the homogeneous equation is found to be proportional to:
$$
\ee^{\nu t}(1-y+y\ee^{-\mu t})^{p\frac{\nu}{\mu}}\;.
$$
Using the initial value $G(z,y,0)=z$,
$$
G^{-1}(z,y,t)=\ee^{\nu t}(1-y+y\ee^{-\mu t})^{p\frac{\nu}{\mu}}\left(\frac{1}{z}+
\nu(p-1)\int_0^t
\ee^{-\nu s}(1-y+y\ee^{-\mu s})^{-p\frac{\nu}{\mu}}\,\dd s\right)\;.
$$
When $\mu=\nu$, the primitive can be explicitly calculated. This is
the only case considered by Bartlett.
$$
\int_0^t \ee^{-\nu s}(1-y+y\ee^{-\nu s})^{-p}\,\dd s
=
\frac{1}{(1-p)\nu y}\left(1-(1-y+y\ee^{-\nu t})^{1-p}\right)\;.
$$
Hence Bartlett's explicit expression:
$$
G^{-1}(z,y,t) = \frac{(1-y+y\ee^{-\nu t})^p}{\ee^{-\nu t}z}+
\frac{(1-y+y\ee^{-\nu t})}{y\ee^{-\nu t}}-
\frac{(1-y+y\ee^{-\nu t})^p}{y\ee^{-\nu t}}\;.
$$
This gives the generating function of both counts, starting with
one normal cell. The generating function of mutant cells
alone is obtained by setting $z=1$ in the expression above. The
generating function of mutant cells, starting with $n_0$ normal
cells is the $n_0$\textsuperscript{th} power:
$$
G^{n_0}(1,y,t)=\left(\frac{(1-y+y\ee^{-\nu t})^p}{\ee^{-\nu t}}+
\frac{(1-y+y\ee^{-\nu t})}{y\ee^{-\nu t}}-
\frac{(1-y+y\ee^{-\nu t})^p}{y\ee^{-\nu t}}\right)^{-n_0}\;.
$$
We want an asymptotic value for this expression, as $p$ tends to $0$,
$n_0$ and $t$ to $+\infty$. Let us replace the two terms in 
$(1-y+y\ee^{-\nu  t})^p$
by their order 1 expansion:
$$
(1-y+y\ee^{-\nu t})^p = 1+p\log(1-y+y\ee^{-\nu t})+o(p)\;.
$$ 
Remember also that $t$ is large, so that:
$$
\log(1-y+y\ee^{-\nu t}) = \log(1-y)+O(\ee^{-\nu t})
$$
One gets:
$$
G^{n_0}(1,y,t)=\left(1+p\frac{y-1}{y\ee^{-\nu
      t}}\log(1-y)+o(p)+pO(\ee^{-\nu t})\right)^{-n_0}
$$
The non-trivial limit is obtained as as $p$ tends to $0$,
$n_0$ and $t$ to $+\infty$, in such a way that 
$$
\lim n_0\,p\,\ee^{\nu t} = \alpha\;,
$$ 
where $\alpha$ is a positive real number. Then:
$$
\lim G^{n_0}(1,y,t) = \exp\left(\alpha \frac{1-y}{y}\log(1-y)\right)
= 
(1-y)^{\alpha\frac{1-y}{y}}\;.
$$
The \emph{Luria-Delbr\"uck} distribution with parameter $\alpha$ is
defined as the probability distribution on integers with generating
function:
$$
g_\alpha(y)=(1-y)^{\alpha\frac{1-y}{y}}\;.
$$
The function $g_\alpha$ has no left derivative at $y=1$: the
Luria-Delbr\"uck distribution has no moment of any order. Denote by
$p_n$ the corresponding probabilities:
$$
g_\alpha(y)=\sum_{m=0}^{+\infty} p_m\,y^m\;.
$$
The first value is obtained for $y=0$: $p_0=\ee^{-\alpha}$.
There is no explicit expression for $p_m$ as a function of $m$. 
An equivalent as $m$ tends to infinity is 
$p_m\sim \frac{\alpha}{m^2}$. The exact
values can be numerically computed through the recursive formula:
$$
p_m=\frac{\alpha}{m}\sum_{i=0}^{m-1} \frac{p_i}{n-i+1}\;.
$$
See Ma \emph{et al.} \cite{Maetal92}, Pakes \cite{Pakes93}, 
and Kemp \cite{Kemp94} for simple 
derivations of the main results
on the $p_n$'s.
As an example, the table below gives some values for the probability
than more than 50 mutant cells remain, for different values of
$\alpha$.
$$
\begin{array}{|c|cccccccccc|}
\hline
\alpha&1&2&3&4&5&6&7&8&9&10\\\hline
\PP[X>50]&0.021&0.045&0.072&0.102&0.136&0.172&0.213&0.256&0.302&0.349\\\hline
\end{array}
$$
As expected from a heavy tail distribution, there are quite
reasonable chances to get sizeable amounts of mutant cells.
\vskip 2mm
How about the Lenski experiment?
As we have seen, the initial number of normal cells in any of the twelve 
10~mL vessels is $5\times
10^6$. Let us take $n_0=6\times 10^7$. For the mutation probability,
Philippe \emph{et al.} \cite{Philippeetal07} give $p=5\times
10^{-10}$ per base pair. This is sensibly lower
than the value given by Kendall \cite{Kendall52}, who simply says 
that the mutation probability is lesser than $10^{-7}$.
As we have seen in section \ref{birth}, $\ee^{\nu t}$ is the expected
number of cells stemming from one initial cell. In the LE, the daily
increase is 100-fold. So we shall retain $\ee^{\nu t}=10^{2}$. The
parameter for the Luria-Delbr\"uck distribution is:
$$
\alpha=n_0\,p\,\ee^{\nu t}=3\;.
$$
For the Luria-Delbr\"uck distribution with parameter $3$, the probability to
get no mutant cells is $p_0=0.05$. The probability to get more than
100 mutant cells is $0.034$.
%
%
\vskip 2mm
The asymptotics of the number of mutants in the general case $\mu\neq
\nu$ is discussed by Jaeger and Sarkar 
\cite{JaegerSarkar95}. To the best of our knowledge, no derivation
from Bartlett's model with $\mu\neq \nu$ has been made. However, we
do not think it would change by much the conclusions. If $\mu>\nu$
(beneficial mutation), mutant cells will multiply faster on average,
(thus be more numerous) than predicted by the Luria-Delbr\"uck
distribution. In other words, the Luria-Delbr\"uck distribution
function is an upper bound for the distribution function of the
number of mutant cells in the general case. The heavy tail property
can only be reinforced.
\vskip 2mm
The Bartlett model that was exposed here is not
unique.  Luria-Delbr\"uck distributions can be obtained through
other models, including the earlier deterministic growth model of
Lea and Coulson \cite{LeaCoulson49} (see Zheng's review
\cite{Zheng99}). It has been extended to other types of non-Markovian
growth models by Dewandji \emph{et al.} \cite{Dewanjietal05}. 
\section{Parameter estimation}
\label{estimation}
Gause \cite{Gause34}, in the introduction to his section ``On the
mechanism of competition in yeast cells'', cites early 30's
publications while making quite clear
statements on the relation between experimental data and mathematical
models. We could not say it better.
\begin{quote}
No mathematical theories can be accepted by biologists without a most
careful experimental verification. We can but agree with the following
remarks made in Nature (H. T. H. P. '31) concerning the mathematical
theory of the struggle for existence developed by Vito Volterra:  ``This
work is connected with Prof. Volterra's researches on
integro-differential equations and their applications to mechanics. In
view of the simplifying hypothesis adopted, the results are not likely
to be accepted by biologists until they have been confirmed
experimentally, but this work has as yet scarcely begun.'' First of
all, very reasonable doubts may arise whether the equations of the
struggle for existence given in the preceding chapter express the
essence of the processes of competition, or whether they are merely
empirical expressions. everybody remembers the attempt to study from a
purely formalistic viewpoint the phenomena of heredity by calculating
the likeness between ancestors and descendants. This method did not
give the means of penetrating into the mechanism of the corresponding
processes and was consequently entirely abandoned. In order to
dissipate these doubts and to show that the above-given equations
actually express the mechanism of competition, we shall now turn to an
experimental analysis of a comparatively simple case. It has been
possible to measure directly the factors regulating the struggle for
existence in this case, and thus to verify some of the mathematical
theories. 

Generally speaking, biologists usually have to deal with empirical
equations. The essence of such equations is admirably expressed in the
following words of Raymond Pearl ('30): ``The worker in practically any
branch of science is more or less frequently confronted with this sort
of problem: he has a series of observations in which there is clear
evidence of a certain orderliness, on the one hand, and evident
fluctuations from that order, on the other hand. What he obviously
wishes to do\ldots{} is to emphasize the orderliness and minimize the
fluctuations about it\ldots{} He would like an expression, exact if
possible, or, failing that, approximate, of the law if there be
one. This means a mathematical expression of the functional relation
between the variables\ldots{}

``It should be made clear at the start that there is, unfortunately, no
methods known to mathematics which will tell anyone in advance of the
trial what is either the correct or even the best mathematical
function with which to graduate a particular set of data. The choice
of the proper mathematical function is essentially, at its very best,
only a combination of good judgment and good luck. In this realm, as
in every other, good judgment depends in the main only upon extensive
experience. What we call good luck in this sort of connection has also
about the same basis. The experienced person in this branch of applied
mathematics knows at a glance what general class of mathematical
expression will take a course, when plotted, on the whole like that
followed by the observations. He furthermore knows that by putting as
many constants into his equation as there are observations in the data
he can make his curve hit all the observed points exactly, but in so
doing will have defeated the very purpose with which he started, which
was to emphasize the law (if any) and minimize the fluctuations,
because actually if he does what has been described he emphasizes the
fluctuations and probably loses completely any chance of discovering a
law. 

``Of mathematical functions involving a small number of constants there
are but relatively few\ldots{}  In short, we live in a world which appears
to be organized in accordance with relatively few and relatively
simple mathematical functions. Which of these one will choose in
starting off to fit empirically a group of observations depends
fundamentally, as has been said, only on good judgment and
experience. There is no higher guide'' (pp. 407-408). 
\end{quote}
The mathematical models that have been presented so far have no
predictive value, until they have been
confronted to experimental data. The parameters should be estimated,
the data have to be adjusted, and the goodness-of-fit must be
tested. The adjustment of population models to bacteria growth
experiments has been the object of countless publications, of which we
have retained a few among the most recent.
\vskip 2mm
Miao \emph{et al.} \cite{Miaoetal11} give an interesting
review of parameter estimation methods for deterministic models.
The general methods for adjusting models are presented by Jaqaman and
Danuser in \cite{Jaqaman06}.
Gutenkunst \emph{et al.} \cite{Gutenkunst07} argue on sloppy parameter
sensitivity in systems biology models.
All models have at least in common growth
rates and mutation probabilities, as parameters to be estimated from
the data. Although most references focus on mutation rates, 
the estimation of growth rates was recently studied by Maruvka \emph{et al.}
\cite{Maruvkaetal11}.
Mutation rates can be estimated by adjusting observed number of mutants in
bacterial growth experiments. The method is usually referred to as
\emph{fluctuation analysis}. It has been presented in many references, 
including Koch \cite{Koch82}, Sarkar \emph{et al.}
\cite{Sarkaretal92}, Stewart \emph{et al.}  
\cite{Stewartetal90}, Jones \emph{et al.} \cite{Jonesetal94}.
Dedicated softwares were made available by Zheng \cite{Zheng02} and
Hall \emph{et al.} \cite{Halletal09}.
Several refinements in the use of the Luria-Delbr\"uck probabilities
have been proposed, in particular by  Kepler \& Oprea
\cite{KeplerOprea01,OpreaKepler01}. 
\vskip 2mm
The estimation of mutation rates was studied as early as 1974
by Crump \& Hoel \cite{CrumpHoel74}. The quality of mutation rates
estimates is discussed by Stewart \cite{Stewart94}.
Different estimation methods are
presented by Foster \cite{Foster06}. Improvements have been proposed
by Koziol \cite{Koziol91}, and by Gerrish  \cite{Gerrish08}.
Grant \emph{et al.} \cite{Grantetal08} made a case study on
\emph{Mycobacterium tubercolosis}. 
\section{Conclusion}
The few classics of mathematical modelling that have been reviewed here
are hoped to be of some potential 
use in the description of different aspects of 
the Lenski experiment. As already said, none of them has any scientific
validity as long as they have not been confronted to real data. 
However, the models share some common qualitative features that are 
summarized here for comparison with actual
observations. Instead of categorizing the models according to
mathematical coherence as we did before, we will summarize their
predictions on four different questions about the LE.
\begin{enumerate}
\item \emph{Day 0: Which number of cells can carry a
  given mutation that was not present the previous day?}\\
All existing models use asymptotics as the initial
  number of cells is large and the probability of mutations is
  low. The random number of new mutants follows a heavy-tailed
  distribution of which the Luria-Delbr\"uck distribution is a
  prototype. This means that with a reasonable probability, sizeable
  amounts of mutants can be present at the end of any given day.
\item \emph{Night 0: What are the chances that a mutation having
  appeared on day 0 disappears through dilution?}\\
Since the daily sampling eliminates each day
  99\% of the cells, mutants present in small quantities have high
  chances to disappear. In particular,
  non-beneficial mutations do not survive. Beneficial mutations
  carried by enough cells (at least a few tens), have good
  chances to be represented the next days.
\item \emph{Day 1: If two different strains of cells are present 
at the beginning of a given day, what will their
  proportion be at the end of the day?}\\
If a beneficial mutation is carried by a certain
  proportion of the population, even a weak one, that proportion will
  gradually increase along the day
with the multiplication of cells. So with
  high probability, the initial proportion will be larger on the next day.
\item \emph{Day N: How many days will it take before a beneficial mutation
  is carried by all cells in the population?}\\
The initial proportion of day $N+1$ is an increasing function of that of the
previous day. The better the fitness of mutants, the steeper that
function. When the initial proportion of mutants is already strong, there
is a high probability that the less fit cells will be wiped out by the
next daily sampling. The invasion of the full
population will take from a few days to a month. 
\end{enumerate}
\bibliographystyle{plain}
\bibliography{IS}
\end{document}